\documentstyle[preprint,tighten,aps,epsfig]{revtex}

\begin{document}
\newcommand{\vn}[1]{\mbox{\boldmath$#1$}} 

\title{{\bf ROTATIONAL DYNAMICS OF THE MAGNETIC PARTICLES IN FERROFLUIDS}
\footnote{Work supported in part by CNPq (Brazil)
 and by Alexander von Humboldt Foundation (Germany).}}

\author{\bf Claudio Scherer$^{(1,2)}$, Hans-Georg Matuttis$^{(1)}$ \\
 (1) {\sl Institute for Computer Applications 1, University of
 Stuttgart}\\ {\sl 70569 Stuttgart, Germany } \\
 (2) {\sl Institute of Physics, Federal University of Rio Grande do Sul}\\
  {\sl 91501-970 Porto Alegre, RS - Brazil}}
 
\date{\today}
\maketitle

\begin{abstract}

 A new theory for the dynamics of the magnetic particles and their 
magnetic moments in ferrofluids is developed. Based on a generalized 
Lagrangian formulation for the equations of motion of the colloidal 
 particle, we introduce its  interaction with the solvent
fluid via  dissipative and  random noise torques, as well as the 
interactions between the particle and its magnetic moment, treated as an 
independent physical entity and characterized by three generalized
coordinates, its two polar angles and its modulus. It has been 
recognized recently that inertial effects, as well as the particle's 
rotational Brownian motion, may play important roles
on the dynamic susceptibility of a class of magnetic fluids. No satisfactory
theory existed, up to now, that takes this effects into account. The theory
presented here is a first-principles 3-dimensional
approach, in contrast to some phenomenological 2-dimensional approaches
that can be found in the recent literature.
It  is appropriate for superparamagnetic, non-superparamagnetic
and mixed magnetic fluids. As a simple application, the blocked limit 
(magnetic moment fixed in the particle) is treated numerically.
The rotational trajectory of the particles in presence of a magnetic
field, as well as the response functions and dynamic susceptibility 
matrices are explicitly calculated for some values of the parameters

\



\end{abstract}

\newpage

\section{Introduction}

Considerable interest has been shown, in recent years, on the 
dynamics of the magnetization of ferrofluids in presence of applied
magnetic fields and on the corresponding complex magnetic
susceptibility. Just to give a few examples of recently published 
work on the field we mention the theoretical works by 
 Raikher and Rusakov\cite{Raikher},  Coffey and Kalmykov\cite{Coffey96},
 Shliomis and Stepanov\cite{Shliomis93}, the experimental works by
Morais et al.\cite{Morais}, Fannin et al.\cite{Fannin96}, Vincent 
et al.\cite{Vincent}  and an experimental-theoretical paper by Fannin 
et al.\cite{Fannin97} . Certainly, this increased interest in a better 
understanding of the behavior of these materials is related to their
renewed technological importance, with various new applications\cite{Raj}.

The usual theoretical approach to calculate the dynamic susceptibility
is based on Gilbert's\cite{Gilbert}  or Landau Lifshitz'\cite{LL}
 equation (which are equivalent) for the dynamics of the magnetic
 moment, with the addition of noise, following the pioneering work of 
Brown\cite{Brown}. Several authors use these equations of motion to 
calculate relaxation times and the susceptibility is then borrowed 
from Debye's theory\cite{Debye}.

Two distinct rotational relaxation mechanisms may coexist in
ferrofluids: the N\'eel relaxation, by which the magnetic moment moves
with respect to the mechanical particle, and the Brownian, or Debye
relaxation, corresponding to the particle's rotation inside the fluid.
In most experimental situations  one of these mechanisms is dominant, 
and this may be the reason why there is not, up to now, that we 
know, a satisfactory theory, sufficiently general to be applied for
all situations, from the pure N\'eel to the pure Brownian relaxation,
passing by all possible combinations of those mechanisms.
In this respect the model of ``two spheres'', by Fannin and
Coffey\cite{Coffey} should be mentioned as a first effort.

The purpose of the present paper is to present such a general
theory. The main limitation of our approach is that we deal only
with axially symmetric particles, with easy axis of magnetization
parallel to the symmetry axis. However, the magnetic moment is allowed
to rotate inside the particle, as well as to have an oscillating
modulus, and the particle is allowed to rotate with respect
to the solvent, which is immobile with respect to the laboratory.
The suspension is considered sufficiently dilute for the 
particle-particle interaction to be negligible, so that we deal
only with single particle dynamics. However, in a mean field approximation,
our approach can serve as a starting point for the inter-particle
interactions to be considered in future works.

In section II we write the equations of the rotational motion of an axially
symmetric particle inside a fluid (Langevin-type equations), based on
the generalized Euler-Lagrange equations. In section III we obtain,
from the equations of section II, in a convenient limit, the
equations of motion for the magnetic moment $\vn{\mu}$, which reduce, 
in the case of constant modulus of $\vn{\mu}$, to the Gilbert's equation. 
In section IV we arrive at the set of six coupled equations, for
the six degrees of freedom, the three Euler angles of the particle's
rotations, the two polar angles of $\vn{\mu}$ and its modulus.
Some less general situations are also considered in this section,
as particular cases. In section V the ``blocked limit'', i.e.,
when the magnetic moment is fixed with respect to the particle,
is treated as an explicit example of application.
In section VI we introduce a simple version of linear response
theory, applicable for the cases where the noise can be considered only
for its effect as a thermal bath. In section VII we apply this
linear response approach to calculate the dynamic susceptibility
of the ferrofluid in the blocked limit and in section VIII some numerical
results are presented and discussed.  
 
We do not explore, in the present paper, the set of six  equations
of section IV, Eqs. (\ref{gem}), in its great generality, because this 
would make the paper too long. Work with this purpose is being
carried out by the authors, to be published in future papers.

\section{Rotational Dynamics of a Particle in a Fluid}

Consider a particle of axially symmetric shape in suspension in a
fluid. The principal moments of inertia will be denoted by
$I_1=I_2$ and $I_3$. Disregarding translational degrees of freedom,
its Lagrangian may be written in terms of the Euler angles
$\theta$, $\phi$ and $\psi$ (in the notation of 
Goldstein\cite{Goldstein}), taken as generalized coordinates, as

\begin{equation}
\label{Lagrangian}
L=\frac{I_1}{2}(\dot{\theta}^2+\dot{\phi}^2 \sin^2\theta)
+\frac{I_3}{2}(\dot{\psi}+\dot{\phi}\cos\theta)^2-V(\theta, \phi)
\end{equation}

\noindent
where $V(\theta, \phi)$ is some orientation dependent potential.
It cannot depend on $\psi$ because of the axial symmetry of the particle.
 
The interaction forces (torques) between the particle and the fluid are of the 
dissipative and noise types. Therefore, they are not included in the
Lagrangian, but instead, we have to use the ``generalized
Euler-Lagrange equations'', with the corresponding torques,
represented by $Q_i$, at the right hand side:

\begin{equation}
\label{GEL}
\frac{d}{dt}\frac{\partial L}{\partial \dot q_i}-
\frac{\partial L}{\partial q_i}=Q_i \; ,
\end{equation}

\noindent where $q_i = \theta, \phi$ or $\psi$.

 We write the non-conservative
torques $Q_i$ as sums of dissipative and noise terms, in the form

\begin{equation}
\label{nct}
Q_i=-\frac{\partial \cal F}{\partial \dot{q_i}}+{\Gamma}_i(t) \; , 
\end{equation}

\noindent where $\cal F$ is the following Rayleigh dissipation function
\cite{Goldstein},

\begin{equation}
\label{ray}
{\cal F}=\frac{1}{2}\lambda ((\dot{\theta}^2+\dot{\phi}^2 \sin^2\theta)+
\frac{1}{2}\lambda'(\dot{\psi}+\dot{\phi}\cos\theta)^2,
\end{equation}

\noindent and ${\Gamma}_i(t)$ are the noise torques. The dissipation
constants $\lambda$ and $\lambda'$ may be different because 
$\lambda'$ is associated with the particle rotation around the
symmetry axis, while  $\lambda$ is associated with the rotations perpendicular
 to it. Substituting Eqs. (\ref{Lagrangian}), (\ref{nct}) and
(\ref{ray}) into Eq. (\ref{GEL}) we obtain the following system of
equations for the particle's rotation:

\begin{mathletters}
\label{em2}
\begin{eqnarray}  
&&I_1(\ddot{\theta}- \dot{\phi}^2 \sin \theta \cos \theta) +
I_3 \; \dot{\phi}\; (\dot{\psi}+\dot{\phi}\cos \theta)\sin \theta +
\lambda \; \dot{\theta}+V_{\theta}={\Gamma}_{\theta}\; ,
 \label{em2a} \\
\nonumber \\
&&I_1( \ddot{\phi} \sin^2 \theta  +2\; \dot{\phi}
\; \dot{\theta} \sin \theta \cos \theta  ) +I_3 \cos \theta \;
 \frac{d}{dt}(\dot{\psi}+\dot{\phi}\cos \theta) + \nonumber \\
&& -I_3 (\dot{\psi}+\dot{\phi}\cos \theta)
 \dot{\theta} \sin \theta
+\lambda \; \dot{\phi} \sin^2 \theta +V_{\phi}={\Gamma}_{\phi} \; ,
\label{em2b} \\
\nonumber \\
&&I_3 \frac{d}{dt}(\dot{\psi}+\dot{\phi}\cos \theta)+
\lambda' \; (\dot{\psi}+\dot{\phi}\cos \theta)={\Gamma}_{\psi}.
\label{em2c}
\end{eqnarray}
\end{mathletters}

\noindent where $V_{\theta}= \partial V / \partial {\theta}$ and
 $V_{\phi}=\partial V / \partial {\phi}$. The expression
 $(\dot{\psi}+\dot{\phi}\cos \theta)$ was left unbroken wherever
it appears in the above equations because it represents the component
of the angular velocity vector ${\vn\omega}$ along the symmetry axis and we 
 make use of this fact in the interpretation of the dissipative 
torques in terms of the components of ${\vn\omega}$, as follows. 

Let us define the
 following four unit vectors: $\vn z$, along the laboratory z-axis, 
 $\vn c$, along the particle's symmetry axis, 
 $\vn a$, perpendicular to the plane 
containing $\vn c$ and  $\vn z$ ($\widehat{\vn c \vn z}$-plane) and
$\vn{b}$, perpendicular to the $\widehat{\vn c \vn a}$-plane, namely,

\begin{mathletters}
\label{uvc}
\begin{eqnarray}
&&\vn{z} = (0,\; 0,\; 1) \; , \label{uvca} \\
&&\vn{c} = (\sin\theta \cos\phi,\; \sin\theta \sin\phi, \;
 \cos\theta)\; , \label{uvcb} \\
&&\vn{a} =\frac{\vn z \times 
\vn c}{\sin\theta}=(-\sin\phi,\cos\phi,0) \; ,\label{uvcc} \\
&&\vn{b}=\vn c \times \vn{a} =(-\cos\theta \cos\phi,\; 
-\cos\theta \sin\phi,\; \sin\theta)\; .\label{uvcd}
\end {eqnarray}
\end{mathletters}

As a notation to be used throughout this work, subscripts  $z, \; c,
 \; a$  or $b$ on a vector indicate its orthogonal projection on the
 $\vn z$, $\vn c$, $\vn a$ or $\vn b$
directions and subscript $\bar c$  indicates the vector's
projection on the plane perpendicular to $\vn c$ . 

 The particle's  angular velocity vector ${\vn\omega}$ may be decomposed 
into a sum of two vectors,  perpendicular and  parallel to  $\vn c$,
 respectively,

\[
{\vn\omega} = {\vn\omega}_{\bar{c}}+{\omega}_{c}\; \vn c \; ,
\]

\noindent with

\[
{\vn\omega}_{\bar{c}}=\vn c \times \dot{\vn c}
= (-\dot{\theta} \sin\phi-\dot{\phi}\sin\theta \cos\theta \cos\phi,
\; \dot{\theta} \cos\phi -\dot{\phi}\sin\theta \cos\theta \sin\phi,
\; \dot{\phi}\sin^2\theta) 
\]

\noindent and

\[
{\omega}_{c} = \dot{\psi}+\dot{\phi} \cos\theta \; .
\]

\noindent The orthogonal projection of ${\vn\omega}_{\bar{c}}$ on the z-axis
is 

\[
\omega_{\bar{c} z}= {\vn\omega}_{\bar{c}}\cdot \vn z=\dot{\phi}
\sin^2\theta \; ,
\]

\noindent and the orthogonal projection of $\vn\omega$ (or of
 ${\vn\omega}_{\bar{c}}$) on the
direction perpendicular to the $\widehat{\vn c \vn z}$ plane is

\[
\omega_ a= {\vn\omega}\cdot \vn a={\vn\omega}_{\bar{c}}\cdot \vn a=
 \dot{\theta} \; .
\]

\noindent Thus we see that the dissipative torques present in 
Eqs. (\ref{em2a}), (\ref{em2b}) and (\ref{em2c})
 are given by $\omega_a$, $\omega_{\bar{c} z}$
and ${\omega}_{c}$, respectively, times the dissipation parameters
$\lambda$ or $\lambda'$. 

The noise torques will be treated along these same lines.
 We start by defining
the noise torque vector by its orthogonal components,

\[
\vn{\Gamma}=\Gamma_a \; \vn a 
+\Gamma_{b} \; \vn{b} + \Gamma_c \; \vn c \; .
\]

The noise becomes completely defined by stating the statistics
of its three components. The usual procedure is to consider them
as statistically independent, Gaussian white noise. This is, however,
not a necessary assumption and we leave it open for future modeling.
What we need now is to know how the three components come into
Eqs. (\ref{em2}). Guided by the above decomposition 
of the dissipative torque, we are led to identify

\begin{eqnarray}
{\Gamma}_{\theta}&=&\Gamma_a  \;, \nonumber \\
{\Gamma}_{\phi}&=&\Gamma_{\bar c z}=\vn{\Gamma}_{\bar c} \cdot \vn z= 
\Gamma_b \; \sin\theta \; ,\nonumber \\
{\Gamma}_{\psi}&=&\Gamma_c \;. \nonumber  
\end{eqnarray}

Before we proceed to deduce the equations of motion for the general
case of magnetic particles in ferrofluids we show, in the next
section, how to obtain, from Eqs. (\ref{em2}), the equations of motion
for the spherical coordinates of a mono-domain magnetic moment.

\section{Equations of Motion for a Magnetic \\ Moment}

 The magnetic moment $\vn \mu$ of a mono-domain particle is related to
 its internal angular momentum $\vn S$ by $ \vn{\mu} = \gamma \vn S$,
where $\gamma$ is the gyromagnetic factor. Although the modulus $ S $ of
$\vn S$ is taken as constant in most works on superparamagnetism
and magnetic fluids, for very small particles its oscillation may
be significant and we prefer to allow it to be time dependent.
The modern technology allows the preparation of samples with
magnetic particles whose diameters are smaller than 
20$\AA$\cite{Chen} and superparamagnetic clusters containing only 12 
magnetic atoms have also been reported\cite{Novak}.
 We can model the magnetic moment by a rotating charged particle,
in the limit of zero moments of inertia, $I_1 \rightarrow 0$,
 $I_3 \rightarrow 0$, and $\dot{\psi} \rightarrow \infty$ so that
$I_3 \dot{\psi}=S$. Because in the next section we will work 
with the joint system, a particle and its fluctuating magnetic
moment, we write the generalized coordinates, potential energy,
dissipative and noise torques, with a notation distinct from that
corresponding to the particle. Namely, we make the following
 substitutions: $ \theta \rightarrow \vartheta, \;\; 
\phi \rightarrow \varphi, \;\; I_3 \, \dot{\psi} \rightarrow S,
\;\; V \rightarrow W,
\;\; \lambda \rightarrow \xi, \; \; \lambda' \rightarrow \xi'$
and $\Gamma \rightarrow \cal T$. We also introduce two modifications
in the equation corresponding to Eq. (\ref{em2c}), namely, we write
$S-S_0$ instead of $S$ in the dissipative term and introduce a torque
$W_s$, whose origin will be explained below.
 In the said limit and with
the new notation the system of Eqs. (\ref{em2}) becomes:

\begin{mathletters}
\label{emm}
\begin{eqnarray}
&& S \; \dot{\varphi} \sin \vartheta +\xi \; \dot{\vartheta}
+W_{\vartheta}={\cal T}_{\vartheta}\; , \label{emma} \\
&&\dot S \cos \vartheta  -S \; \dot{\vartheta} \sin \vartheta +
\xi \; \dot{\varphi} \sin^2 \vartheta +W_{\varphi}=
{\cal T}_{\varphi} \; ,\label{emmb} \\
&&\dot S+\xi' (S-S_0)+W_s={\cal T}_s.\label{emmc}
\end{eqnarray}
\end{mathletters}

\noindent Here we have written $S-S_0$, instead of $S$, in the dissipation
 term of Eq. (\ref{emmc})
 to account for the fact that the relaxation of 
the fluctuations of $S$ is towards a most probable (equilibrium)
value $S_0$ and not towards 0. It may appear strange that, even though
we have derived the equations of motion for $\vn S$ from the equations
of motion for  a symmetric particle, in a convenient limit, we have now
to add a term ``ad hoc'' ($S_0$), which does not have an equivalent 
in the particle's equations. This is so because in classical physics
the equilibrium magnetization is always zero. Non-zero equilibrium
magnetic moments can only exist because of the quantum mechanical nature 
of matter and, therefore, cannot be deduced from a pure classical
approach. The torque $W_s$ was introduced because  a crystal field
may have an effective interaction  with $\vn{\mu}$, with origin
in an orbital contribution to $\vn S$\cite{White}, with a
 possible torque component parallel to $\vn S$. There is not an
 equivalent term in Eq. (\ref{em2c}) because of the assumed axial 
symmetry of the particle. 

It is interesting to study the behavior of Eqs. (\ref{emm}) in the
absence of noise, ${\cal T}_i=0$ and with $W_s=0$.
 Eq. (\ref{emmc}) has then the
trivial stationary solution $S=S_0$. Assuming this constant value
for $S$ in Eqs. (\ref{emma}) and (\ref{emmb}) they reduce to

\begin{mathletters}
\label{emm0}
\begin{eqnarray}
&& S_0 \; \dot{\varphi} \sin \vartheta +\xi \; \dot{\vartheta}
+W_{\vartheta}=0 \;,\label{emm0a} \\
&& -S_0 \; \dot{\vartheta} \sin \vartheta +
\xi \; \dot{\varphi} \sin^2 \vartheta +W_{\varphi}=0 \; .\label{emm0b}
\end{eqnarray}
\end{mathletters}

The conservative torques, $-W_{\vartheta}$ and $-W_{\varphi}$, have, 
usually, contributions from two different origins, the interaction
of $\vn S$ with a crystalline, anisotropy field and/or with a
 magnetic field, which can also be of several different origins.
 In the case of magnetic field, $\vn{H}$, the potential energy is
$W=-\vn{\mu} \cdot \vn{H}$. With a little of algebraic work one can 
show, in this case, that the set of Eqs. (\ref{emm0}) is equivalent
to the well known Gilbert's equation\cite{Gilbert},

\[
\frac{d \vn{\mu}}{dt}=\gamma \; \vn{\mu} \times \left[ \vn{H}-
\frac{\xi}{\mu^2}\frac{d \vn{\mu}}{d t} \right]\; ,
\]

\noindent for $\vn{\mu}= \gamma \vn{S}$ and $S=S_0$. This equation was
used by W. F. Brown\cite{Brown} as a starting point for his stochastic theory
of superparamagnetism, where he assumed the magnetic field $\vn{H}$ to contain
a noise term. A more general theory for superparamagnetism,
 which allows also for oscillations on the modulus $\mu=\gamma S$ of
 the magnetic moment, was worked out by Ricci and Scherer
\cite{Tri1,Tri2,Tri3}, based on the set of Eqs. (\ref{emm}). For
this reason we will not continue to explore the consequences of
Eqs. (\ref{emm}) in the present paper, turning, instead, to the more
general ferrofluid, where the rotations of the mechanical particle are
taken into account, in addition to the motion of $\vn S$ relative to
the particle.

\section{The General Ferrofluid}

In recent years several researchers\cite{Raikher,Morais,Coffey,Fannin}
 have drawn attention to the
importance of the motion of the magnetic particle, its inertia and 
viscous interaction with the fluid, to the dynamic magnetic
susceptibility of ferrofluids. A theoretical treatment of this problem,
which is both, more fundamental and more general than those previously
published, follows naturally from the context described above.
 
Taken together, the systems of Eqs. (\ref{em2}) and (\ref{emm}) contain 
all the degrees of freedom relevant to the problem.
To the potential energy terms, $V$ in Eqs. (\ref{em2}) and $W$ in
Eqs. (\ref{emm}), the interaction energy
between the magnetic moment and the particle, which we will denote by
 $U $, has to be added.
Due to the particle's symmetry, this term can only depend  on $S$ and on
the angle between $\vn S$ and the symmetry axis, $\vn c$.
It is convenient to define another  orthogonal set of unit vectors,
related to the direction of the magnetic moment, namely,
 $\vn s$, in the $\vn S$ direction, $\vn u$, perpendicular
to the $\widehat{\vn s \vn z}$-plane and $\vn v$, perpendicular to the
 $\widehat{\vn s \vn u}$-plane, 

\begin{mathletters}
\label{uvs}
\begin{eqnarray}
&&\vn s=\frac{\vn S}{S}=
 (\sin\vartheta \cos\varphi,\; \sin\vartheta \sin\varphi,\;
 \cos\vartheta )\;, \label{uvsa} \\
&&\vn u=\frac{\vn z \times \vn s}{\sin\vartheta}=
(-\sin\varphi,\cos\varphi,0) \; ,\label{uvsb} \\
&&\vn v=\vn s \times \vn u=(-\cos\vartheta \cos\varphi,\; 
-\cos\vartheta \sin\varphi,\; \sin\vartheta)\; .\label{uvsc} 
\end{eqnarray}
\end{mathletters}

\noindent The interaction energy $U$ can then be written as
 $U(S, \vn s \cdot \vn c)$. 
 In principle the particle can interact also with other fields,
 besides $\vn H$, as is the case if it has an electric dipole and an electric
field is present. For this reason we keep also the potential energy
$V(\theta,\phi)$  in the new set of equations.

The dissipative interaction associated with the rotation of $\vn S$
relative to the particle  will be written in terms of the relative
angular velocity vector.
 Since only rotations perpendicular to $\vn S$ can lead to a meaningful 
interaction torque with origin on the relative motion, we define the
relative angular velocity $\vn{\omega}_r$ as

\[
\vn{\omega}_r=\vn{\varpi}-\vn{\omega}_{\bar{s}} \; , 
\]

\noindent where

\[
\vn{\varpi}=\vn s \times \dot{\vn s}
\]

\noindent is the angular velocity of rotation of the magnetic moment with
respect to the laboratory and

\[
\vn{\omega}_{\bar{s}}=\vn s \times \vn{\omega} \times \vn s 
= \vn{\omega}-(\vn s \cdot \vn{\omega}) \vn s \; .
\]

\noindent is the orthogonal projection of the particle's angular velocity
${\vn\omega}$ on the plane perpendicular to $\vn S$.
The dissipative interaction torque on the particle is then
$+\xi \; \vn{\omega}_r$. The plus sign is because 
of the way we defined $\vn{\omega}_r$, where the particle's angular
velocity appears with a minus sign.
Guided by the interpretation of the dissipative torque terms of
Eqs. (\ref{em2}) in terms of angular velocity components, as explained
bellow the said equations, we write down immediately the dissipative 
torque terms to be added to the left-hand sides (therefore, with
a $-$ sign) of Eqs. (\ref{em2}), namely
 
\begin{eqnarray}
&&-\xi \, \omega_{ra}=-\xi \, \vn{\omega}_r \cdot \vn a \;,\nonumber \\
&&-\xi \, \omega_{r\bar{c}z}=-\xi \,  
[\vn{\omega_r}-(\vn{\omega_r}\cdot \vn c)\, \vn c ] \cdot \vn z 
=-\xi \, (\omega_{rz}-\omega_{rc}\, \cos\theta \;,\nonumber \\
&&-\xi \, \omega_{rc}=-\xi \; \vn{\omega}_r \cdot \vn c \;. \nonumber
\end{eqnarray}

\noindent Of course, all this scalar products, as well as those which
follow, in the next equations, may be easily written as functions
of the four angles $\theta, \; \phi, \; \vartheta$ and $\varphi$ 
and their time derivatives, by using Eqs. (\ref{uvc}) and (\ref{uvs}).
However, because scalar products are very easily handled in numerical
procedures, we prefer to leave them in this form.

 Clearly, the torque on the magnetic moment, 
due to the relative motion, is the ``reaction'' to the torque on the 
particle, i.e., it is  equal to  $-\xi \vn{\omega}_r$, and, in place of
 $\xi \; \dot{\vartheta}$ and  $\xi \; \dot{\varphi} \sin^2 \vartheta$ 
in Eqs. (\ref{emm}) we shall use (remembering that 
$\vn{\omega}_{r\bar{s}}=\vn{\omega}_r$)

\begin{eqnarray}
&&\xi \, \omega_{ru} = \xi \, \vn{\omega}_r \cdot \vn u \; ,\nonumber \\
&&\xi \, \omega_{rz} = \xi \, \vn{\omega}_r \cdot \vn z \; .\nonumber
\end{eqnarray}

\noindent No term coming from the relative angular
velocity $\vn{\omega}_r$ has to be added to  Eq. (\ref{emmc})
 because $\vn{\omega}_r$ is perpendicular to $\vn S$.
However, there is the term $\xi' \; (S-S_0)$ already  present 
in that equation, with origin in the (quantum)
fluctuations of $S$, and this term will be kept. Since angular 
momentum has to be conserved, its reaction counterpart on the
particle has to be added to Eqs. (\ref{em2}). Calling

\[
\vn{\cal R}= (S-S_0)\; \vn s\; ,
\]

\noindent the terms to be added to the left-hand sides of  Eqs. (\ref{em2})
are

\begin{eqnarray}
 &&-\xi' \;{\cal R}_{a}=-\xi' \;\vn{\cal R} \cdot \vn a 
=-\xi' \; (S-S_0) \, \vn s \cdot \vn a\; ,\nonumber \\
 &&-\xi' \;{\cal R}_{\bar c z}=-\xi' \;
[\vn{\cal R} - (\vn{\cal R} \cdot \vn c )\, \vn c ]\cdot \vn z
=-\xi' \; (S-S_0)[\vn s - (\vn s \cdot \vn c )\, \vn c ]\cdot \vn z 
\; ,\nonumber  \\
 &&-\xi' \;{\cal R}_{c}=-\xi' \;\vn{\cal R} \cdot \vn c 
=-\xi' \; (S-S_0)\, \vn s \cdot \vn c \; .\nonumber 
\end{eqnarray}

The noise torques of interaction between the particle and the 
magnetic moment can be written down along the same lines of 
procedure as done for the noise torques of the fluid on the particle,
at the end of section II. We assume three orthogonal,
independent, noise torque vectors, along the unit vectors
defined with respect to the direction of the magnetic moment:

\begin{equation}
\label{TR}
\vn{\cal T}={\cal T}_{s}\; \vn s + {\cal T}_{u}\; \vn u
+{\cal T}_{v}\; \vn v \; .
\end{equation}
 
\noindent Being $\vn{\cal T}$ the torque on the magnetic moment, then
the torque on the particle is $-\vn{\cal T}$. Following the
same line of reasoning as done before, we identify the torques in
Eqs. (\ref{emm}):

\begin{eqnarray}
&&{\cal T}_{\vartheta}= {\cal T}_u \; , \nonumber \\
&&{\cal T}_{\varphi}={\cal T}_{\bar s z} ={\cal T}_v 
\sin\vartheta \; , \nonumber \\
&&{\cal T}_s={\cal T}_s \; .\nonumber
\end{eqnarray}

\noindent Correspondingly, we the following terms have to be added
to the  right-hand-sides of Eqs. (\ref{em2}):

\begin{eqnarray}
&&{\cal T}_{\theta}=-{\cal T}_a=-\vn{\cal T} \cdot \vn a \; ,
 \nonumber \\
&&{\cal T}_{\phi}=-{\cal T}_{\bar{c}z}=-[\vn{\cal T}-
( \vn{\cal T}\cdot \vn c )\, \vn c ] \cdot \vn z
 \; ,\nonumber \\
&&{\cal T}_{\psi}=-{\cal T}_c= -\vn{\cal T} \cdot \vn c \; . \nonumber 
\end{eqnarray} 

Therefore, the state of the composed system, the particle and its magnetic
moment, is described by the 6 generalized coordinates,
$\theta, \; \phi, \; \psi, \vartheta, \; \varphi$ and $S$, whose
dynamical behavior is governed by the following set of
coupled differential equations:

{\samepage
\begin{mathletters}
\label{gem}
\begin{eqnarray}  
&&I_1(\ddot{\theta}- \dot{\phi}^2 \sin \theta \cos \theta) +
I_3 \; \dot{\phi}\; (\dot{\psi}+\dot{\phi}\cos \theta)\sin \theta +
\lambda \; \dot{\theta}-\xi {\omega}_{ra}+ 
 \nonumber \\
&&-\xi' \;{\cal R}_a+V_{\theta}+U_{\theta}=
{\Gamma}_a-{\cal T}_a\; , \label{gema} \\ 
\nonumber \\
&&I_1( \ddot{\phi} \sin^2 \theta  +2\; \dot{\phi}
\; \dot{\theta} \sin \theta \cos \theta  ) +I_3 \cos \theta \;
 \frac{d}{dt}(\dot{\psi}+\dot{\phi}\cos \theta) + \nonumber \\
&& -I_3 (\dot{\psi}+\dot{\phi}\cos \theta)
 \dot{\theta} \sin \theta
+\lambda \; \dot{\phi} \sin^2 \theta -\xi {\omega}_{r \bar c z}
-\xi' \; {\cal R}_{\bar c z} +V_{\phi}
+U_{\phi}={\Gamma}_b \sin\theta -{\cal T}_{\bar c z}\; , \label{gemb}\\
\nonumber \\
&&I_3 \frac{d}{dt}(\dot{\psi}+\dot{\phi}\cos \theta)+
\lambda' \; (\dot{\psi}+\dot{\phi}\cos \theta)-\xi {\omega}_{rc}
-\xi' \; {\cal R}_c =
{\Gamma}_c -{\cal T}_c \; ,\label{gemc} \\
\nonumber \\
&& S \; \dot{\varphi} \sin\vartheta + \xi {\omega}_{ru}
+W_{\vartheta}+U_{\vartheta}=+{\cal T}_u\; , \label{gemd} \\
\nonumber \\
&&\dot S \cos \vartheta  -S \; \dot{\vartheta} \sin \vartheta +
 \xi {\omega}_{r \bar c z}+W_{\varphi}+U_{\varphi}=
+{\cal T}_{\bar s z} \; ,\label{geme} \\
\nonumber \\
&&\dot S+\xi' (S-S_0)+U_S={\cal T}_s \; .\label{gemf}
\end{eqnarray}
\end{mathletters}
}
This set of six equations
is of very general applicability on ferrofluids. It allows for a large
variety of modeling: There are three independent conservative
interaction potentials, $V, \; U$ and $W$,  four
 dissipative parameters, $\lambda, \; \lambda', \; \xi$, and
$\xi'$, and also the noise torques $\vn{\Gamma}$ and $\vn{\cal T}$,
whose statistical properties are open for modeling. Particle-particle
interaction was not explicitly taken into account, but on a mean-field
approximation it can be included in $V$ and/or in $W$. 

As we mentioned before, in most cases of practical interest
the fluctuations in the modulus $S$ can be neglected. In this
case Eqs. (\ref{gem}) become simpler, in several respects:
Eq. (\ref{gemf}) ceases to exist, all terms in $\dot S$ and in
$\xi'$ become zero and the noise term ${\cal T}_{s}$ in Eq. (\ref{TR})
and its contribution in Eqs. (\ref{gem}) also vanish.

Two interesting limit situations are readily obtained from
Eqs(\ref{gem}), the ``superparamagnetic'' limit, for which the
particle's coordinates, $\theta, \; \phi$, and $\psi$, are taken as 
constants, so that the system reduces to the last three equations,
and the ``blocked'' limit (also called ``Brownian''
limit\cite{Shliomis} or ``inertial limit''\cite{Coffey96}),
 when the magnetic moment is blocked along the particle's symmetry 
direction, i.e.,  $ \vartheta=\theta$ and $\varphi= \phi$, but the 
particle can rotate  inside the fluid. The superparamagnetic limit
has been treated in three previous papers by Ricci and 
Scherer\cite{Tri1,Tri2,Tri3} and also by other authors. In the 
remaining sections of this paper we will deal with the ``blocked'' limit.

\section{Dynamics of the Magnetic Moment in the Blocked Limit}
 
We consider now the situation in which the magnetic moment is blocked
along the particle's symmetry axis. This may happen because the sample
is kept below the ``blocking temperature'' $T_B$ \cite{TB} or because
the material is so highly anisotropic that the magnetic moments
only exists parallel to the easy axis\cite{Novak}. The particle
is still immersed in a fluid carrier, being able to rotate, 
together with its magnetic moment.

In terms of the set of Eqs. (\ref{gem}), the blocked limit is obtained
by assuming an interaction potential $U$ of the form
$-U_0 \delta(\vn s - \vn c)$, with $U_0 \rightarrow \infty$, so that 
the only states energetically possible are those with $\vn s=\vn c$,
i.e., $\vartheta=\theta$ and $\varphi=\phi$. By summing
 Eq. (\ref{gema}) with  Eq. (\ref{gemd}) and  Eq. (\ref{gemb})
 with Eq. (\ref{geme}) 
the interaction terms $U_{\theta}$ and $U_{\vartheta}$ as well as
 $U_{\phi}$ and $U_{\varphi}$
 cancel out. The terms containing $\omega_{ra}$, $\omega_{r \bar c z}$,
 ${\cal R}_a$,
 ${\cal R}_{\bar c z}$, $\vn{\cal T}_a$, and $\vn{\cal T}_{\bar c z}$
 become identically zero, and 
${\cal R}_c$ becomes $(S-S_0)$. Choosing $\theta$ and $\phi$
to denote the common polar angles, the system of equations becomes:

\begin{mathletters}
\label{gembl}
\begin{eqnarray}
&&I_1(\ddot{\theta}- \dot{\phi}^2 \sin \theta \cos \theta) +
I_3 \; \dot{\phi}\; (\dot{\psi}+\dot{\phi}\cos \theta)\sin \theta +
\lambda \; \dot{\theta} +V_{\theta} +
\nonumber \\&&+ S \; \dot{\phi} \sin\theta +W_{\theta} = {\Gamma}_a\; ,
 \label{gembla} \\ 
\nonumber \\
&&I_1( \ddot{\phi} \sin^2 \theta  +2\; \dot{\phi}
\; \dot{\theta} \sin \theta \cos \theta  ) +I_3 \cos \theta \;
 \frac{d}{dt}(\dot{\psi}+\dot{\phi}\cos \theta) + \nonumber \\
&& -I_3 (\dot{\psi}+\dot{\phi}\cos \theta)
 \dot{\theta} \sin \theta
+\lambda \; \dot{\phi} \sin^2 \theta +V_{\phi}+\dot S \cos\theta 
\;+ \nonumber \\
&&-S \; \dot{\theta} \sin\theta +W_{\phi}={\Gamma}_b \sin\theta \; ,
\label{gemblb} \\
\nonumber \\
&&I_3 \frac{d}{dt}(\dot{\psi}+\dot{\phi}\cos \theta)+
\lambda' \; (\dot{\psi}+\dot{\phi}\cos \theta)
- \xi' (S-S_0)={\Gamma}_c-{\cal T}_c \; ,\label{gemblc} \\
\nonumber \\
&&\dot S+\xi' (S-S_0)=+{\cal T}_c \; .\label{gembld}
\end{eqnarray}
\end{mathletters}

 We will not explore, in the present paper, the system of
 Eqs. (\ref{gembl}) in all its generality. In the following we will
consider only the cases when the noise terms do not need to be taken
into account explicitly. The explicit 
presence of white noise torques makes of Eqs. (\ref{gem}) and
 (\ref{gembl}) Ito-Langevin systems and their treatment demands the
 use of stochastic calculus. This will be treated in a future paper.
Several circumstances may be devised where neglecting the noise is a
 good approximation. One of them is when the system is drawn far from
equilibrium, in presence of a strong magnetic field, with 
$\mu   H \gg k_B T$. Then the relaxation to the new
 equilibrium state, with $\vn{\mu}$ approximately parallel to $\vn H$,
proceeds without an important influence of the noise. The Bloch's
 equations of magnetic resonance are based on this idea: they contain
relaxation terms (with relaxation times $T_1$ and $T_2$), but no noise
terms. Another interesting circumstance is in the context of linear
response theory. The usual formulation considers the noise
sufficiently weak for its effect to be only in establishing an initial
thermal equilibrium. The perturbing field is then introduced
 adiabatically, i.e., with the system disconnected from the thermal bath.
A similar approach to linear response, however based on the equations
of motion, instead of based on the Liouville equation for the 
probability distribution, which 
is the case of Kubo's linear response theory\cite{Kubo}, will be 
presented in the next section.

In the following we will also assume a constant  modulus for the
magnetic moment, i.e., $S=S_0$, and for the interaction potential
we consider only $W=-\vn{\mu} \cdot \vn{H}=-\gamma S_0  
\vn s \cdot \vn{H}$. For simplicity, only a constant
field,  $\vn{H}=H_0 \vn{z}$, will be considered now,
 but in the section on linear response we
will introduce also a time dependent transversal field.
 
With this simplifications, the system of Eqs. (\ref{gembl}) becomes

{\samepage
\begin{mathletters}
\label{gemnn}
\begin{eqnarray}
&&I_1(\ddot{\theta}- \dot{\phi}^2 \sin \theta \cos \theta) +
I_3 \; \dot{\phi}\; (\dot{\psi}+\dot{\phi}\cos \theta)\sin \theta +
\lambda \; \dot{\theta} +
\nonumber \\&&+ S_0 \; \dot{\phi} \sin\theta +
 \gamma S_0 H_0 \sin\theta  =0\; ,
 \label{gemnna} \\ 
\nonumber \\
&&I_1( \ddot{\phi} \sin^2 \theta  +2\; \dot{\phi}
\; \dot{\theta} \sin \theta \cos \theta  ) +I_3 \cos \theta \;
 \frac{d}{dt}(\dot{\psi}+\dot{\phi}\cos \theta) + \nonumber \\
&& -I_3 (\dot{\psi}+\dot{\phi}\cos \theta)
 \dot{\theta} \sin \theta
+\lambda \; \dot{\phi} \sin^2 \theta -S_0 \; \dot{\theta} \sin\theta=0
\label{gemnnb} \\
\nonumber \\
&&I_3 \frac{d}{dt}(\dot{\psi}+\dot{\phi}\cos \theta)+
\lambda' \; (\dot{\psi}+\dot{\phi}\cos \theta)=0 \; . \label{gemnnc}
\end{eqnarray}
\end{mathletters}
}

 Eq. (\ref{gemnnc}) shows that, under the circumstances
considered, and for any value that the function
 $\dot{\psi}+\dot{\phi}\cos \theta$
may have, due to initial conditions, it will relax exponentially to zero.
Since we will consider, in what follows,  only stationary initial
conditions, its value will be taken as identically zero. This 
simplifies the above  set of equations to

\begin{mathletters}
\label{emnn}
\begin{eqnarray}
&&I_1(\ddot{\theta}- \dot{\phi}^2 \sin \theta \cos \theta) +
 S_0 \; \dot{\phi} \sin\theta +\lambda \; \dot{\theta} =
- \gamma S_0 H_0 \sin\theta \; , \label{emnna}  \\ 
&&I_1( \ddot{\phi} \sin^2 \theta  +2\; \dot{\phi}
\; \dot{\theta} \sin \theta \cos \theta  ) 
 -S_0 \; \dot{\theta} \sin\theta+\lambda \; \dot{\phi} \sin^2 \theta=0
\label{emnnb} \; .
\end{eqnarray}
\end{mathletters} 

This equations may be solved numerically, for arbitrarily given
parameters, $I_1, S_0, \lambda$ and $\gamma H_0$ and arbitrary 
initial conditions, by using standard methods. An example of
solution is shown in Fig.1, in form of a rotational trajectory of the
magnetic moment, drawn over a sphere to help visualization. The origin
of the magnetic moment vector is at the center of the sphere and its
head follows the trajectory on the surface. the magnetic field
$\vn{H_0}$ is parallel to the line from  the south to the  north pole.
For Fig.1-a the initial velocities $v_0=\dot{\theta}(t_0)$ and 
$w_0=\dot{\phi}(t_0)$ were arbitrarily chosen, for Figs.1-b and 1-c
the initial velocities were calculated from Eqs. (\ref{w0}) and
(\ref{v0}) of Sec.VIII. 
 The dissipation parameter $\lambda$
for Fig.1-c is 100 times the value used for Figs.1-a and 1-b. All
other parameters and also the total time interval are the same for the
three figures. 

\begin{center}
\epsfig{file=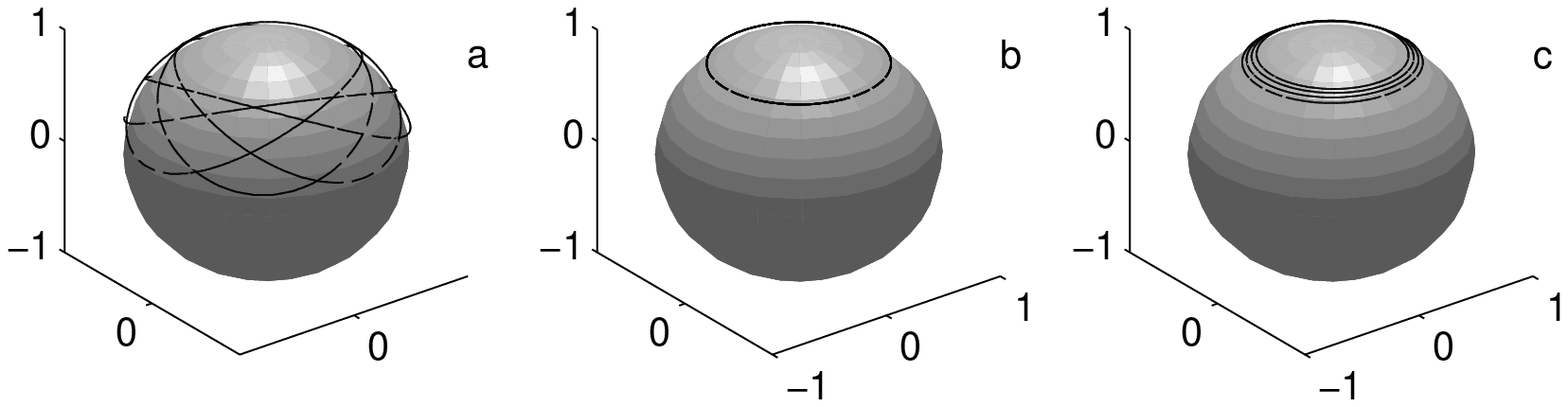,width=15cm}
\end{center}

{\small
\noindent Fig.1: Trajectory of the head of the magnetic moment vector
in its rotation around the constant magnetic field (see text).
 1-a: for arbitrarily chosen initial velocities;
1-b: for initial velocities given according to Sec. VIII;
1-c: same as 1-b, but with dissipation parameter, $\lambda$,
100 times larger.}

\section{A Simple Approach to Linear Response \\ Theory}

The standard Linear Response Theory\cite{Kubo} is based on the
classical or quantum Liouville Equation, for the probability
distribution function or the density matrix, respectively, for
systems in absence of noise, or based on the
Fokker-Planck Equation in the case of stochastic processes\cite{Tri3}.
A simpler approach, which has some limitations, but is appropriate for
the purposes of this work, based on the equations of motion of the
system, is as follows.

\subsection{The Linear Equations for the Perturbation}

Let us assume that the system's coordinates satisfy the equation

\begin{equation}
\label{first}
\vn{\dot Q}(t)=\vn P (\vn{Q},t) \; ,
\end{equation}

\noindent where $\vn{Q}=(q_1, q_2, \cdots, q_n)$ and where
 $\vn{P}=(P_1, P_2, \cdots, P_n)$ are functions of the
 coordinates and of the time. We assume further that the only
explicit time dependence of  $\vn{P}$ comes from an applied
perturbing field,

\[
\vn F(t)=(F_1(t), \cdots, F_m(t)) \; ,
\]

\noindent in the form

\begin{equation}
\label{AF}
\vn P(\vn Q,t)=\vn{P}^0(\vn Q) + {\widetilde A(\vn Q)}\cdot \vn F(t) \; ,
\end{equation}

\noindent where $\widetilde A$ is an $n \times m$ matrix.
 The perturbation $\vn F(t)$
is assumed sufficiently weak for its effect in deviating
$\vn Q(t)$ from its unperturbed values to be well approximated by a
a linear functional. This approach is exact in the limit $\vn F
\rightarrow 0$, which defines the initial susceptibility.
In this equation and in what follows 
an upper index $^0$ (like in $\vn{P}^0$) will indicate
``unperturbed values'', while a lower index $_0$ (like in $\vn{Q}_0$)
 will be used for initial values. Therefore $\vn{Q}^0(t)$ indicates
the solution of the unperturbed equation, with given initial
conditions, 

\begin{equation}
\label{UP}
\vn{\dot Q}^0(t)=\vn P^0 (\vn{Q}^0) \; , \;\;\;\;\;\;\;\ 
\vn Q^0(t_0)=\vn Q_0^0=(q_{01}, \cdots, q_{0n}) \; .
\end{equation}

The solution of Eq. (\ref{first}) will be written as 

\begin{equation}
\label{solf}
\vn Q(t)=\vn Q^0(t)+\vn X(t)\; ,
\end{equation}

\noindent where $\vn X(t)$ is a linear functional of $\vn F$. Substituting
Eq. (\ref{solf}) into Eq. (\ref{first}), using Eq. (\ref{AF}) and keeping
only linear terms in $\vn F$ follows

\begin{equation}
\label{ex}
\vn{\dot X}(t)=\widetilde K(\vn Q^0(t))\cdot \vn X(t)
+ \widetilde A(\vn Q^0(t)) \cdot \vn F(t)\;,
\end{equation}

\noindent where the matrix elements of $\widetilde K$
are

\begin{equation}
\label{mel}
K_{ij}=\frac{\partial P^0_i}{\partial q^0_j}\;.
\end{equation}

Eq. (\ref{ex}) is a linear, inhomogeneous, system of equations
for $\vn X$, with time dependent coefficients $ K_{ij}(t)=
 K_{ij}(\vn Q^0(t))$. The corresponding homogeneous equation,

\begin{equation}
\label{hom}
\vn{\dot Y}=\widetilde K \cdot \vn Y \; 
\end{equation}

\noindent has the solution

\[
\vn Y(t)=\widetilde M(t,t_0)\cdot \vn Y_0 \; ,
\]

\noindent where the matrix $\widetilde M$ is formally given by

\begin{equation}
\label{M}
\widetilde M(t,t_0)=\exp(\int_{t_0}^t \widetilde K(t') \, dt')\; .
\end{equation}

\noindent The general solution of Eq. (\ref{ex}) may be written formally as

\[
\vn X(t)=\widetilde M(t,t_0)\cdot \vn X_0 +
\int_{t_0}^t \widetilde M(t,t')\cdot \widetilde A(t')\cdot \vn F(t')\, dt' \; .
\]

\noindent Since $\vn X(t)$ has to be a linear
functional of $ \vn F$, it follows that $\vn X_0=0$. Thus

\begin{equation}
\label{sollin}
\vn X(t)=\int_{t_0}^t \widetilde M(t,t')\cdot
 \widetilde A(t')\cdot \vn F(t')\, dt' \; .
\end{equation}

In numerical procedures, it is often simpler to solve Eq. (\ref{hom})
then to work with Eq. (\ref{M}) to obtain the matrix elements of
$\widetilde M$. To obtain $M_{ij}(t,t_0)$ from the solutions of 
Eq. (\ref{hom})
we define the set of $n$ unit vectors of dimensionality $n$, 

\begin{eqnarray}
\vn Y_0^1=(1,0,\cdots, 0) \nonumber \\
\vn Y_0^2=(0,1,\cdots, 0) \nonumber \\
\cdots \cdots \cdots \cdots \cdots \nonumber \\
\vn Y_0^n=(0,0,\cdots, 1) \nonumber 
\end{eqnarray}

The solution of Eq. (\ref{hom}), with initial condition
$\vn Y(t_0)=\vn Y_0^i$ will be denoted by $\vn Y^i(t)$, i.e.,

\[
\vn Y^i(t)=\widetilde M(t,t_0)\cdot \vn Y_0^i \; .
\]

\noindent Therefore, the $j^{th}$ component of  $\vn Y^i(t)$ is 

\begin{equation}
\label{YijT}
 Y_j^i(t)= M_{ji}(t,t_0) \; ,
\end{equation}

\noindent from what follows that to obtain the matrix elements
$M_{ji}(t,t_0)$  one solves the linear
set of Eqs. (\ref{hom}) with the $\vn Y_0^i$  as initial vectors.

\subsection{Response Function and Susceptibility Matrices}

Consider the dynamical variable (observable)
 $\vn B(\vn Q) = (B_1, \cdots , B_m)$. Its
ensemble average over the initial equilibrium distribution will be 
denoted by $\left< \vn B(\vn Q) \right>_0$. The ``response'' of
$\vn B$ to the perturbing field $\vn F(t)$ is defined by

\[
\delta \vn B(t) \equiv \left< \vn B(\vn Q(t)) \right>_0
-\left< \vn B(\vn Q^0(t)) \right>_0 \; .
\]

Expanding the first term above around $\vn Q^0$ to first order in $\vn X $
and using Eq. (\ref{sollin}) follows

\begin{equation}
\label{response}
\delta \vn B(t)=\left<\widetilde{\nabla B}(t)\cdot \int_{t_0}^t
\widetilde M(t,t')\cdot \widetilde A(t')\cdot \vn F(t') \, dt'
\right>_0 \; ,
\end{equation}

\noindent where $\widetilde{\nabla B}$ is an $m \times n$ matrix,
 with elements

\begin{equation}
\label{nabb}
(\nabla B)_{ki} \equiv B_{k,i}\equiv 
\left( \frac{\partial B_k}{\partial q_i} \right)_
{Q= Q^0(t)} \; , \;\;\; k=1 \cdots  m \; , \;\;  i=1 \cdots n \; . 
\end{equation}

The ``response function matrix'' ($m \times m$) $\widetilde{\Phi}$
 is defined by

\begin{equation}
\label{respfun}
\delta \vn B(t)=\int_{t_0}^t \widetilde{\Phi}(t-t') \cdot \vn F(t')\,
dt'\; .
\end{equation}

\noindent Therefore, comparing Eqs. (\ref{respfun}) and (\ref{response}),
we identify

\begin{equation}
\label{rf}
 \widetilde{\Phi}(t-t')=
\left< \widetilde{\nabla B}(t) \cdot 
\widetilde M(t,t') \cdot \widetilde A(t') \right>_0 \; ,
\end{equation}

\noindent which is  function of $t-t'$ and not of $t$ and $t'$, 
independently, because the average is over the equilibrium 
distribution, for which absolute times are meaningless. Therefore,
we can choose $t'=t_0=0$ in Eq. (\ref{rf}) and write
the matrix elements of $\widetilde{\Phi}(t)$ as

\begin{equation}
\label{melrf}
{\Phi}_{kl}(t)=\sum_{ij}\left< B_{k,i}(t)M_{ij}(t,0)A_{jl}(0)
\right>_0 \; .
\end{equation}

The complex susceptibility is defined as the Fourier-Laplace
transform of  $\widetilde{\Phi}$,

\begin{equation}
\label{chi}
{\chi}_{kl}(\omega)=\lim_{\epsilon \rightarrow 0^+}\int_0^{\infty}
\exp(\rm{i}\omega t' - \epsilon t') \Phi_{kl}(t')\, dt' \; .
\end{equation}

In the next section we apply the concepts and results presented above
to the ferrofluid in the blocked limit.

\section{Dynamical Susceptibility of the Blocked \\ Ferrofluid}

The starting point to apply the linear response approach of last
section is the set of equations of motion for the system. We rewrite
Eqs. (\ref{emnn}) in the form of Eq. (\ref{first}), by defining the 
variables $v=\dot{\theta}$ and $w=\dot{\phi}$. To simplify the notation 
we also introduce $\bar S=S_0/I_1, \; \bar{\lambda}=\lambda /I_1$
 and write Eqs. (\ref{emnn}) in the form

\begin{mathletters}
\label{lin0}
\begin{eqnarray}
&&\dot{\theta}=v \; , \label{lin0a} \\
&&\dot{\phi}=w \; , \label{lin0b} \\
&&\dot{v}= w^2 \sin\theta \cos\theta - \bar S \, w \sin\theta
-\bar{\lambda}\, v - \bar S \, \gamma \, H_0 \sin\theta \; , 
\label{lin0c} \\
&&\dot{w}= -2\, v\, w \cot\theta + \bar S \, v \,\csc\theta
- \bar{\lambda} \, w \; . \label{lin0d}
\end{eqnarray}
\end{mathletters}

By comparison of Eqs. (\ref{lin0}) with Eq. (\ref{UP}) we see that
the components of the vector $\vn P^0(\vn Q^0)$ are just the 
RHS's of Eqs. (\ref{lin0}). From Eq. (\ref{mel}) we obtain the matrix 
elements $K_{ij}(t),\; i,j=\theta,\, \phi, \, v, \, w$. The only non
zero elements are

\begin{eqnarray}
&&K_{\theta v}=1 \; , \nonumber \\
&&K_{\phi w}=1 \; , \nonumber \\
&&K_{v \theta}=w^2(1-2\sin^2\theta)-
\bar S(\gamma \, H_0 + w)\cos\theta \; , \nonumber \\
&&K_{vv}=-\bar{\lambda} \; , \nonumber \\
&&K_{vw}=(2\,w\cos\theta-\bar S)\sin\theta \; , \nonumber \\
&&K_{w \theta}=(2\, v \, w - \bar S \, v \cos\theta) / \sin^2\theta 
 \; , \nonumber \\
&&K_{wv}=(-2\, w \cos\theta + \bar S )/ \sin\theta \; , \nonumber \\
&&K_{ww}=-2\, v \cot\theta - \bar{\lambda} \; ,  \nonumber 
\end{eqnarray}

\noindent where $\theta = \theta(t)$, $v=v(t)$ and $w=w(t)$ are the
solutions of Eqs. (\ref{lin0}). For any given set of initial values
$\theta_0$, $v_0$ and $w_0$ it corresponds a set of 
functions $K_{ij}(t)$  (independent of  $\phi_0$) and, from 
Eqs,(\ref{hom}) and (\ref{YijT}) follows the corresponding
 $Y_i^j(t)$ and $M_{ij}(t)$.

We assume now that a time-dependent perturbing magnetic field is
applied perpendicular to the z-axis, i.e., 
$\vn H(t)=(H_x(t), H_y(t), 0)$. The interaction energy of this field
with the particles magnetic moment is

\[
W=-\vn{\mu} \cdot \vn H=-S_0 \, \gamma \, H_x \, \sin\theta
\cos\phi - S_0 \, \gamma \, H_y \, \sin\theta \sin\phi \; .
\]

\noindent The terms to be added to Eqs. (\ref{lin0c}) and (\ref{lin0d})
 are

\begin{eqnarray}
&&\frac{-1}{I_1}\frac{\partial W}{\partial \theta}=
\gamma \bar S \cos\theta (H_x \cos\phi + H_y \sin\phi ) \; ,\nonumber \\ 
&&\frac{-1}{I_1 \sin^2\theta} \frac{\partial W}{\partial \phi}=
\frac{\gamma \bar S}{\sin\theta}(-H_x \sin\phi + H_y \cos\phi) \; . 
\nonumber  
\end{eqnarray}

\noindent  By adding this terms to the RHS of
Eqs. (\ref{lin0c}) and (\ref{lin0d}), respectively, comparing
with Eqs. (\ref{first}) and (\ref{AF}), and identifying
$F$ with $H$, we can write down the matrix 
$\widetilde A$:

\begin{equation}
\label{mA}
\widetilde A =
\left(
\begin{array}{cc}  
A_{\theta x} & A_{\theta y} \\
A_{\phi x}   & A_{\phi y}  \\
A_{v x}      & A_{v y}     \\
A_{w x}      & A_{w y} 
\end{array}
\right)
=
\left(
\begin{array}{cc}  
0                              & 0            \\
0                              & 0           \\
\gamma\bar S \cos\theta \cos\phi     &\;\;\gamma \bar S \cos\theta \sin\phi  \\
- \gamma\bar S \sin\phi / \sin\theta &\;\;\gamma \bar S \cos\phi/\sin\theta
\end{array}
\right)
\end{equation}

\noindent To calculate the response functions ${\Phi}_{kl}(t)$ by
Eq. (\ref{melrf}) only the matrix elements at time zero are needed.
For this purpose we substitute $\theta$ and $\phi$ in Eq. (\ref{mA})
by $\theta_0$ and $\phi_0$. 
 
 To complete the argument of the average in Eq. (\ref{melrf}) 
we still need the $B_{k,i}(t)$. As the observable vector $\vn B$
we choose the projection of the magnetic moment $\vn \mu$ on the 
$\widehat{xy}$-plane, i.e., $\vn B=(B_x,\; B_y)$, with

\begin{eqnarray}
&&B_x=\gamma \, S_0 \sin\theta \, \cos\phi \; , \nonumber \\
&&B_y=\gamma \, S_0 \sin\theta \, \sin\phi \; . \nonumber
\end{eqnarray}
 
\noindent From Eq. (\ref{nabb}) we then get

\begin{mathletters}
\label{Bki}
\begin{eqnarray}
&&B_{x, \theta}=\gamma \, S_0 \cos\theta \, \cos\phi \; , \label{Bkia}\\
&&B_{x, \phi}=-\gamma \, S_0 \sin\theta \, \sin\phi \; , \label{Bkib}\\
&&B_{y, \theta}=\gamma \, S_0 \cos\theta \, \sin\phi \; , \label{Bkic}\\
&&B_{y, \phi}=\gamma \, S_0 \sin\theta \, \cos\phi \; , \label{Bkid}
\end{eqnarray}
\end{mathletters}

\noindent and all other $B_{k,i}$ are zero.

For the response function $\Phi_{xx}(t)$ we obtain, from
Eq. (\ref{melrf}) 

\begin{equation}
\label{fixx}
\Phi_{xx}(t)=\left< B_{x,\theta}M_{\theta  v}A_{vx} 
+ B_{x,\theta}M_{\theta  w}A_{wx} 
+ B_{x,\phi}M_{\phi  v}A_{vx} 
+ B_{x,\phi}M_{\phi  w}A_{wx} \right>_0 \; ,
\end{equation}

\noindent with time arguments $B_{x,i}(t),\; M_{ij}(t,0)$
and $A_{jx}(0)$.
Substituting the $B_{x,i}$ and $A_{jx}$  by their values as given by
Eqs. ({\ref{Bki}) and (\ref{mA}), writing $\phi(t)=\phi_0 +\Delta\phi$,
where $\Delta\phi$ is independent of $\phi_0$ (because the unperturbed
equations do not contain $\phi$) and using some trigonometric
relations, the average over $\phi_0$ in Eq. (\ref{fixx}) may be done
analytically, resulting in

\begin{eqnarray}
\label{phixx}
\Phi_{xx}(t)= \left< \frac{\gamma^2\,S_0 \bar S}{2} 
\left[ \cos\theta \, \cos\theta_0
\cos\Delta\phi \; M_{\theta v} +\frac{\cos\theta}{\sin\theta_0}
\sin\Delta\phi \; M_{\theta w}+ \right. \right. \nonumber \\
\left.\left. -\sin\theta \, \cos\theta_0 \, \sin\Delta\phi \; M_{\phi v}
+\frac{\sin\theta}{\sin\theta_0} \cos\Delta\phi \; M_{\phi w}\right]
\right>_0 \; .
\end{eqnarray}

By the same procedure we obtain the other response functions:

\begin{eqnarray}
\label{phixy}
\Phi_{xy}(t)=-\Phi_{yx}(t)=
 \left<\frac{\gamma^2\,S_0 \bar S}{2}\left[ \cos\theta \, \cos\theta_0
\cos\Delta\phi \; M_{\theta v} +\frac{\cos\theta}{\sin\theta_0}
\sin\Delta\phi \; M_{\theta w}+ \right.\right. \nonumber \\
\left.\left.-\sin\theta \, \cos\theta_0 \, \sin\Delta\phi \; M_{\phi v}
+\frac{\sin\theta}{\sin\theta_0} \cos\Delta\phi \; M_{\phi w}\right]
\right>_0 \; .
\end{eqnarray}

\noindent Eq. (\ref{phixy}) also shows that Onsager's relation

\[
\Phi_{yx}(t,-H_0)=\Phi_{xy}(t,H_0) \; ,
\]

\noindent is satisfied, because $\Phi_{xy}$ is odd under $H_0 \rightarrow - H_0$.
We also obtain $\Phi_{yy}=\Phi_{xx}$, which is an obvious result,
from symmetry.

\section{Numerical Results and Conclusions}

The equilibrium average indicated in Eqs. (\ref{phixx}) and
(\ref{phixy}) are, in principle, to be done over $\theta_0,\;
v_0$, $w_0$ and all particle's independent parameters (polydispersity).
 The average over $\phi_0$ was already performed
analytically. However, the deviations of $v$ and $w$ from their most 
probable values are rapid fluctuations due to the Brownian noise, 
which is not present in our system of equations. Therefore, the
choice of initial distribution, to be consistent with our
approximation of neglecting the noise, should not include this
fluctuations. We will use Boltzmann's equilibrium distribution
for $\theta_0$ and, for every selected value of $\theta_0$, we
chose $v_0$ and $w_0$ from an approximate solution
of Eqs. (\ref{lin0c}) and (\ref{lin0d}) around $t=0$, calculated in
the following way:

\

\noindent 1) Assume that if $v$ is not zero  at $t= 0$, when we disconnect the
noise, it relaxes to zero according to
the equation $\dot v = -\bar{\lambda}\, v$. 
Eq. (\ref{lin0c}) then leads to

\begin{equation}
\label{w0}
w_0=\frac{\bar S - \sqrt{\bar S^2+4\gamma H_0 \bar S \cos\theta_0}}
{2 \cos\theta_0}\; ,
\end{equation}

\noindent where we have chosen the $-$ sign because $w_0$ has to vanish
 for $H_0=0$.

\

\noindent 2) Analogously, we assume that if $w$ is not $w_0$,
it relaxes to $w_0$ according to the equation 
$\dot w =-\bar{\lambda}\,(w-w_0)$. Eq. (\ref{lin0d}) then leads to

\begin{equation}
\label{v0}
v_0=\frac{\bar{\lambda}\,w_0}{\bar S-2\,w_0\,cos\theta_0}\; .
\end{equation}

\noindent This prescription was used in Sec.V  for the initial velocities
in the  calculation which led to Figs. 1-b and 1-c.
 We remark that this choice of $w_0$ is meaningful only if

\begin{equation}
\label{condition}
\bar S+4\gamma H_0 \cos\theta_0 \geq 0\; .
\end{equation}

\noindent However, if $\gamma\, S_0\, H_0/k_B T$ is not too small,
the Boltzmann distribution becomes negligible for values of
$\theta_0$ such that Eq. (\ref{condition}) is not satisfied.

It is also important to examine under which circumstances the
approximation made in Eq. (\ref{gemnn}), and in all that follows
 that equation, the neglecting of the noise terms, is appropriate. 
We are specially interested in obtaining the dynamical susceptibility
of the system, and therefore we will examine the consequences of that
 approximation in the context of linear response.
Two characteristic times are of importance: The longitudinal
relaxation time  $T_1 \approx \langle \dot\theta \rangle^{-1}$,
where the average is over all magnetic particles, and the transverse
relaxation time $T_2$, which is the time taken by the response
functions to become approximately zero. We have borrowed the notation
$T_1$ and $T_2$ from Magnetic Resonance (MR) because of the similarity of
their meanings here and in NMR or EPR. The longitudinal (parallel to $\vn H_0$)
relaxation, called ``spin-lattice relaxation'' in MR, occurs in a time 
$T_1$, via energy loss to the environment, due to the dissipative torque. 
The transverse relaxation, characterized by the vanishing of
$\Phi_{xx}(t)$ and called ``spin-spin relaxation'' in MR, occurs in a
time $T_2$, due to  the loss of phase coherence between the responses
of the different particles to a pulse of the perturbing field at $t=0$,
 in their rotations around $\vn H_0$. Since we assume
that the system is very close to thermal equilibrium, at a given
temperature, and since the longitudinal relaxation takes the system
out of the initial equilibrium distribution corresponding to that
temperature, just because the thermal noise is neglected in the
equations of motion, our calculation of the response functions and
 susceptibility is only a good approximation if $T_2 \ll T_1$.

There are several sources of transverse relaxation. One of them is,
of course, the noise, which is being neglected in this approximation.
The different initial angles $\theta_0$ and different particle's
parameters $I_1,\; \lambda,$ and $S_0$ (polydispersity) lead to
different frequencies (see Eqs. (\ref{w0}) and (\ref{v0})) and, 
consequently, to loss of phase coherence and to transverse relaxation.
These are taken into account in the averaging procedure on 
Eqs. (\ref{phixx}) and (\ref{phixy}).

To obtain the functions $\theta(t),\; \Delta\phi(t)$ and $M_{ij}(t)$
we have to solve the systems of differential equations,
Eqs. (\ref{hom}) and (\ref{lin0}). We have done it with the Runge-Kutta
algorithm, in a work-station. The particle's parameters, field
intensity and time unit  have been arbitrarily chosen  and kept the
same in all calculations whose results are shown in the figures,
except where explicitly stated.

Polydispersity was treated for particles made of the same material
 and  having the same shape, assuming a uniform distribution
of a linear dimension, $r$,  in an interval of size $\Delta r$,
 i.e., $r$ was chosen to be uniformly distributed in the interval 
$(1-\frac{1}{2}\Delta r,\; 1+\frac{1}{2}\Delta r)$. 
The other particle's parameters were scaled accordingly, namely, 

\begin{eqnarray*}
&& S_0 \propto r^3, \;\;\; I_1 \propto r^5,  \\
&& \bar S = S_0/I_1 \propto r^{-2}, \\
&& \lambda \propto r^3, \;\;\; \bar{\lambda}=\lambda / I_1 \propto r^{-2}.
\end{eqnarray*}

The average over the initial angle $\theta_0$ was weighted
with the Boltzmann equilibrium distribution,

\[
P(\theta_0) \propto \sin\theta_0 \, 
\exp(S_0 \, \gamma \, H_0 \, \cos\theta_0 / k_B T )\; ,
\]

\noindent and the temperature was chosen so that
$S_0 \, \gamma \, H_0 \, \cos\theta_0 / k_B T \approx 5,$
so that $P(\theta_0)$ has a maximum around $\pi /6$.

\begin{minipage}{7 cm}
\begin{center}
\epsfig{file=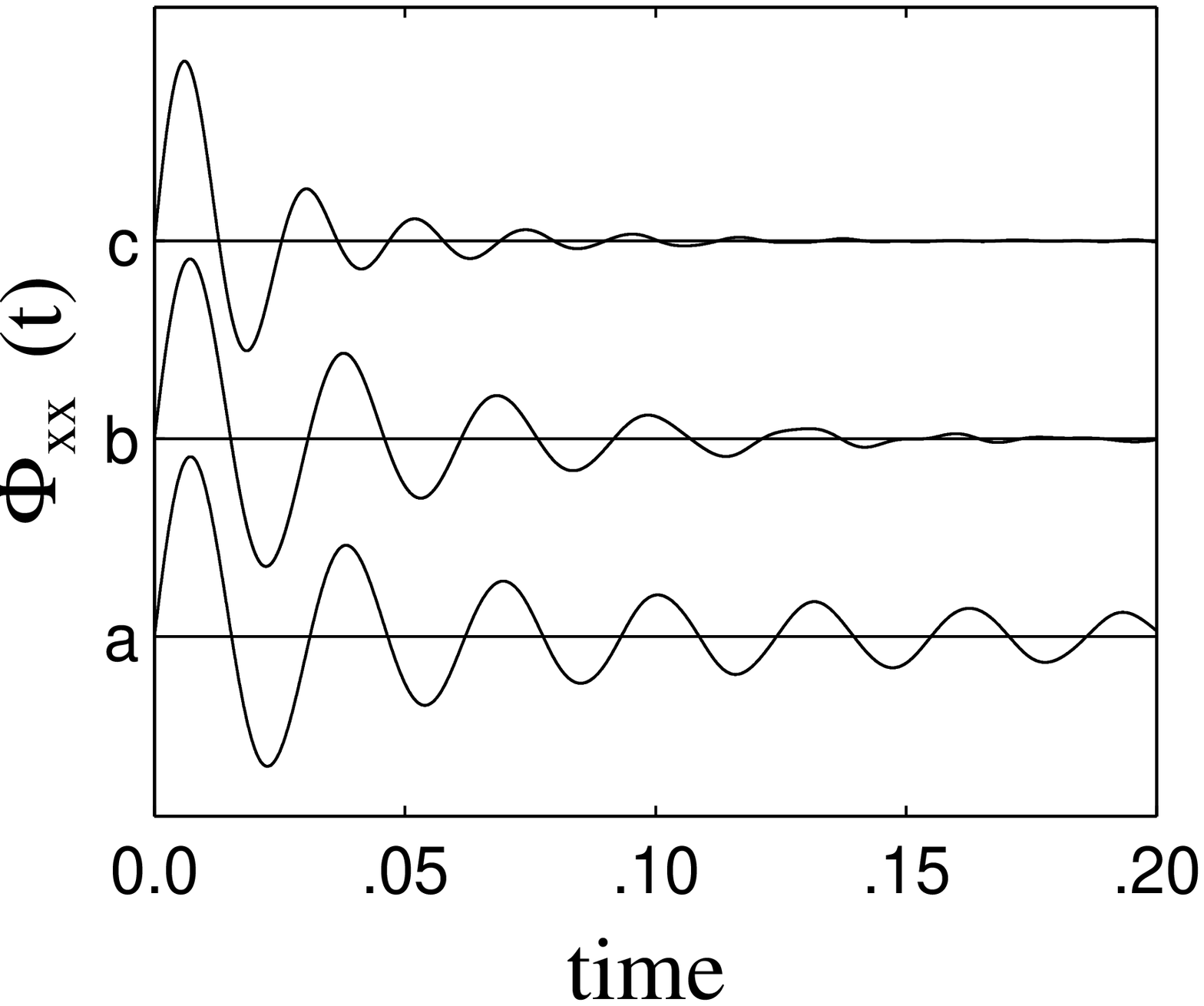,width=5.5cm,angle=-90}
\end{center}
{\small
\noindent Fig.2:  $\Phi_{xx}(t)$ versus time. Line a:  $\Delta r = 0$;
Line b: $\Delta r = 0.05$; Line c: $\Delta r = 0.20$.
}
\end{minipage}
\hspace{1 cm}
\begin{minipage}{7 cm}
Fig.2 shows the diagonal response function $\Phi_{xx}(t)$ versus time,
for three different polydispersity ranges, $\Delta r = 0, \; 0.05$ and
$0.20$, for the curves a, b and c, respectively. This figure confirms
what we said above that polydispersity causes the relaxation time
$T_2$ to decrease. Their values, for the curves a, b and c, 
may be estimated to be approximately $T_2 \approx 0.2, \; 0.1$ and
$0.05$, respectively. 
The longitudinal relaxation time $T_1$ cannot
be estimated from this curves, but has to be calculated together
with the numerical solution of Eqs. (\ref{lin0}), by using
 $T_1 \approx \langle \dot\theta \rangle^{-1}$. 
\end{minipage}

The result is not
strongly dependent on polydispersity, giving, in the present case,
$T_1 \approx 20$. Therefore the condition stated above for the
 appropriateness of the approximation of neglecting the explicit 
presence of noise in the
equations of motion, $T_2 \ll T_1$, is amply satisfied for the
parameters used here.

 Fig.3 and Fig.4 show
the real and imaginary parts of the susceptibility
$\chi (\omega)$, respectively,
corresponding to the response functions of Fig.2. Among other 
information, Fig.4 shows clearly that the broader the dispersity of
particle's size, the broader also the resonance line, as one should
expect. 

\begin{minipage}{7 cm}
\begin{center}
\epsfig{file=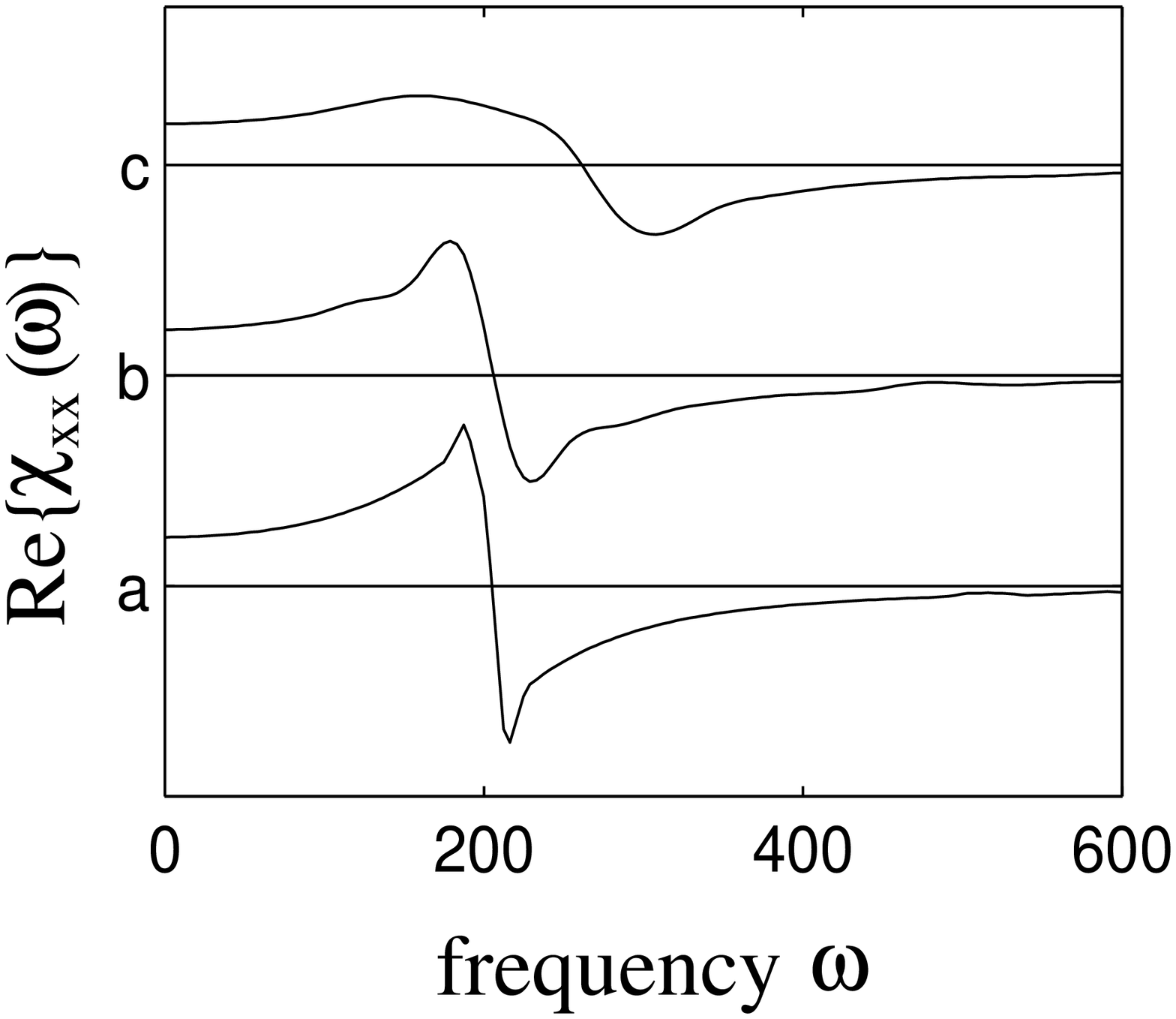,width=6cm,angle=-90}
\end{center}
{\small
\noindent Fig.3: Real part of the susceptybility $\chi 
(\omega)$, versus $\omega$,
corresponding to the response functions of Fig.2.   
}
\end{minipage}
\hspace{1 cm}
\begin{minipage}{7 cm}
\begin{center}
\epsfig{file=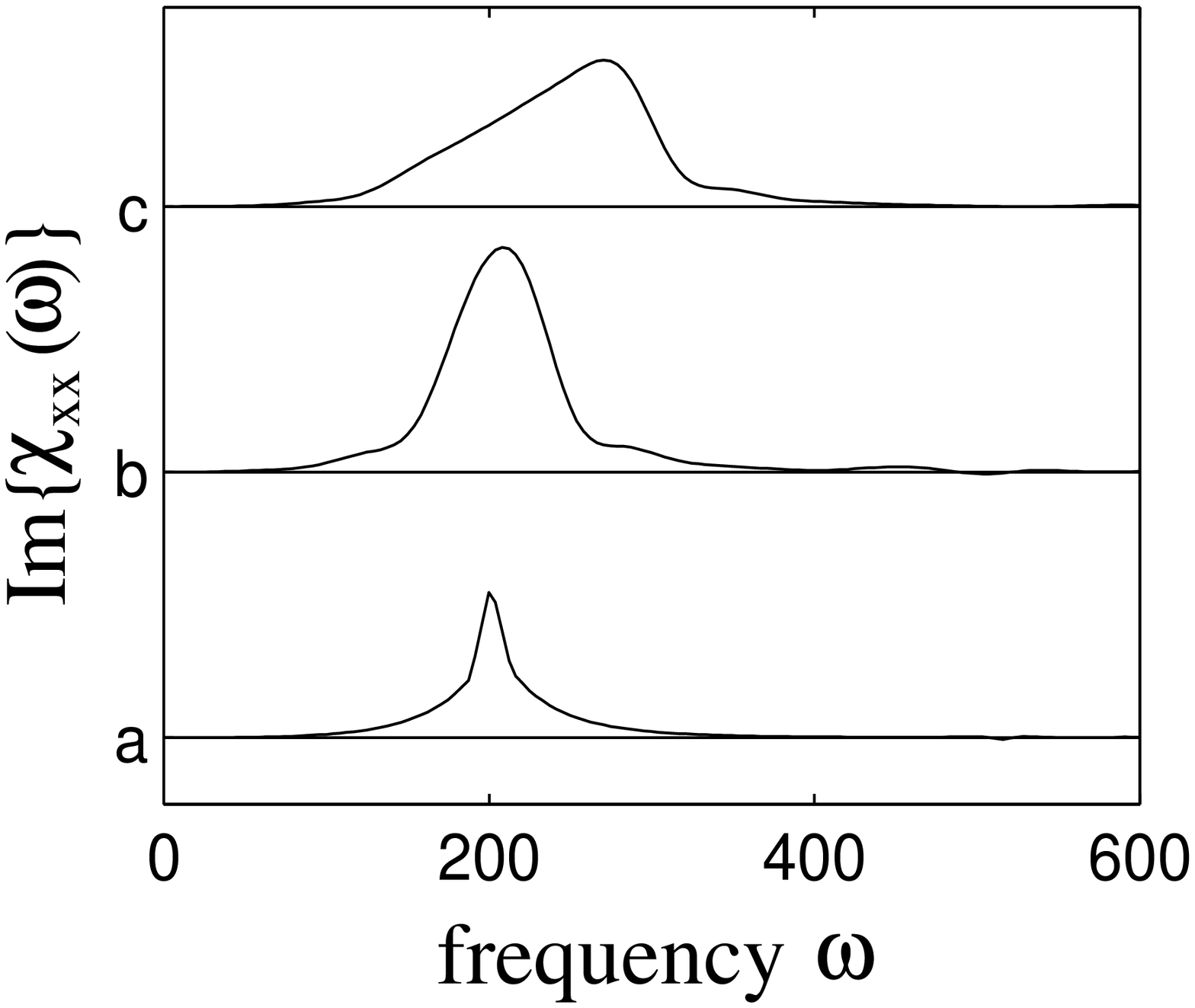,width=6cm,angle=-90}
\end{center}
{\small
\noindent Fig.4: Imaginary part 
of the susceptybility $\chi (\omega)$, versus $\omega$,
corresponding to the response functions of Fig.2.
}
\end{minipage}

\bigskip

\begin{minipage}{7 cm}
\begin{center}
\epsfig{file=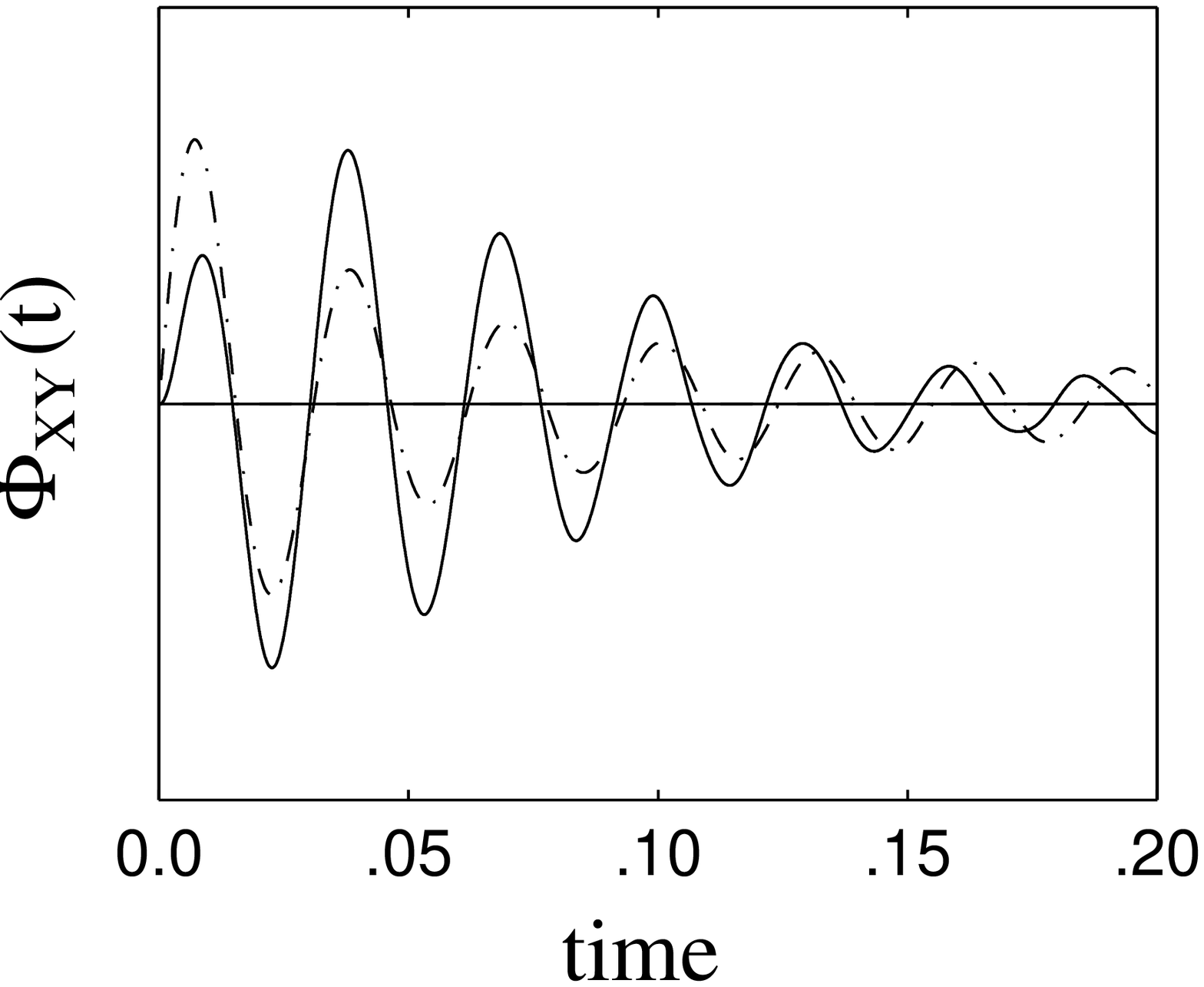,width=6cm,angle=-90}
\end{center}
{\small
\noindent Fig.5: Solid line: $\Phi_{xy}(t)$; 
dot-dash line: $\Phi_{xx}(t)$, the same as in Fig.2-a.
}
\end{minipage}
\hspace{1 cm}
\begin{minipage}{7 cm}
\begin{center}
\epsfig{file=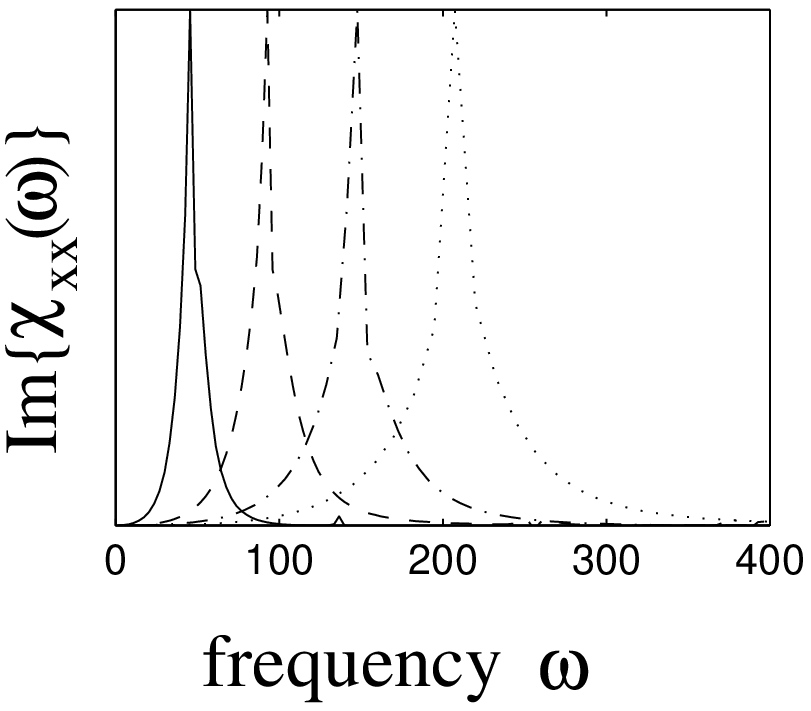,width=6cm,angle=-90}
\end{center}
{\small
\noindent Fig.6: Imaginary part of the suceptibility for 
different values of the moment of inercia, $I_1$. 
Full line: $I_1=1.0$; dashed line: $I_1=0.25$;
dot-dash line: $I_1=0.10$; dotted line: $I_1=0.05$.
}
\end{minipage}

\newpage

Fig.5 shows the non diagonal response function $\Phi_{xy}(t)$ (solid
line), for a monodisperse sample, plotted together with $\Phi_{xx}(t)$ 
(dot-dash line) for comparison. In Fig.6 we compare the resonance
frequency (center of the resonance peak in the imaginary part
of the susceptibility), for different values of the
moment of inertia, $I_1$, keeping constant all other parameters. The 
lowest value, $I_1=0.05$ (dotted line) is the same as used in the 
previous figures, for monodisperse samples. The other curves correspond
to $I_1=0.10$ (dot-dash line), 0.25 (dashed line) and 1.0 (full
line). Clearly, the heavier the particles, the lower the resonance
frequency.

\bigskip

{\bf Acknowledgments:} We thank Dr. Sieghard Weinketz for a critical
reading of the manuscript and good suggestions and one of us (C.S.) 
thanks Prof. Hans Herrmann, director of ICA1, for the nice hospitality
in his institute.

\end{document}